\documentclass[conference]{IEEEtran}
\IEEEoverridecommandlockouts
\usepackage{cite}
\usepackage{amsmath,amssymb,amsfonts}
\usepackage{algorithmic}
\usepackage{graphicx}
\usepackage{textcomp}
\usepackage{xcolor}
\usepackage{amsmath}  
\usepackage{amssymb}  
\usepackage{multirow}
\usepackage{array}   
\usepackage{bm} 
\usepackage{siunitx} 
\usepackage{graphicx}
\usepackage{subcaption}
\usepackage{afterpage}
\usepackage{float} 
\usepackage{caption}
\def\BibTeX{{\rm B\kern-.05em{\sc i\kern-.025em b}\kern-.08em
    T\kern-.1667em\lower.7ex\hbox{E}\kern-.125emX}}
\begin{document}

\title{Adaptive AUV Hunting Policy with Covert
Communication via Diffusion Model\\
}

\author{
    \IEEEauthorblockN{
        Xu Guo\IEEEauthorrefmark{1}, 
        Xiangwang Hou\IEEEauthorrefmark{2}, 
        Minrui Xu\IEEEauthorrefmark{3},
        Jianrui Chen\IEEEauthorrefmark{4},
        Jingjing Wang\IEEEauthorrefmark{4},
        Jun Du\IEEEauthorrefmark{2},
        and Yong Ren\IEEEauthorrefmark{2}
    }    
    \IEEEauthorblockA{\IEEEauthorrefmark{1}Tsinghua Shenzhen International Graduate School, Tsinghua University, Shenzhen, 518055, China \\
    \IEEEauthorrefmark{2}Department of Electronic Engineering, Tsinghua University, Beijing, 10084, China \\
    \IEEEauthorrefmark{3}School of Computer Science and Engineering, Nanyang Technological University, Singapore, 639798, Singapore \\
    \IEEEauthorrefmark{4}School of Cyber Science and Technology, Beihang University, Beijing 100191, China}
    Email: guo-x24@mails.tsinghua.edu.cn, xiangwanghou@163.com,
    minrui001@e.ntu.edu.sg, 
    \\chen\_jr@buaa.edu.cn,
    drwangjj@buaa.edu.cn,
    jundu@tsinghua.edu.cn,
    reny@tsinghua.edu.cn }
\vspace{-9cm}
\maketitle

\begin{abstract}
Collaborative underwater target hunting, facilitated by multiple autonomous underwater vehicles (AUVs), plays a significant role in various domains, especially military missions. Existing research predominantly focuses on designing efficient and high-success-rate hunting policy, particularly addressing the target's evasion capabilities. However, in real-world scenarios, the target can not only adjust its evasion policy based on its observations and predictions but also possess eavesdropping capabilities. If communication among hunter AUVs, such as hunting policy exchanges, is intercepted by the target, it can adapt its escape policy accordingly, significantly reducing the success rate of the hunting mission.
To address this challenge, we propose a covert communication-guaranteed collaborative target hunting framework, which ensures efficient hunting in complex underwater environments while defending against the target's eavesdropping. To the best of our knowledge, this is the first study to incorporate the confidentiality of inter-agent communication into the design of target hunting policy. Furthermore, given the complexity of coordinating multiple AUVs in dynamic and unpredictable environments, we propose an adaptive multi-agent diffusion policy (AMADP), which incorporates the strong generative ability of diffusion models into the multi-agent reinforcement learning (MARL) algorithm. Experimental results demonstrate that AMADP achieves faster convergence and higher hunting success rates while maintaining covertness constraints. 
\end{abstract}


\section{Introduction}
Multiple autonomous underwater vehicles (AUVs) enabled collaborative target hunting has been widely used in  various fields, particularly military missions. However, collaborative underwater hunting tasks present numerous challenges including tracking, obstacle avoidance, and formation control\cite{10272682}. This complexity has garnered significant academic attention, prompting extensive research in the field.

At the early stages, some works\cite{wang2017limit}  \cite{deghat2014localization} explore the problem of capturing stationary or slowly moving targets. These studies typically assume that the targets lack the ability to obtain information on the hunters' positions or velocities. However, these assumptions are highly idealized. Real-world AUVs often possess advanced intelligence, enabling them to detect the hunters' positions using sonar and other sensors, and to respond with evasion policy accordingly. As a further development, the studies \cite{10298247} and \cite{10243453} consider a more practical scenario, where they assume that in the pursuit-evasion interaction between hunter AUVs and the target, the target could observe hunters' locations. In fact, targets not only have observational abilities but also possess eavesdropping capabilities in actual scenarios. If sensitive information (e.g., collaborative policy, locations) exchanged within the formation of hunter AUVs is intercepted by the targets, they will adjust their evasion policy accordingly, which significantly impacts the success rate of the hunting process. However, existing research overlooks the impact of information leakage on the hunting process. To address this, we introduce covert communication techniques into the hunting framework. Covert communication \cite{10090449} involves creating uncertainty in transmissions to prevent adversaries from intercepting sensitive information. Accordingly, we propose a covert communication-guaranteed collaborative target hunting framework that facilitates effective coordination among hunter AUVs while adhering to covert communication constraints, thereby preventing hunting policy from being compromised.

Moreover, coordinating multiple AUVs is an exceedingly complex task. Traditional rule-based methods for AUV target hunting \cite{MENG2021108268} \cite{9069245} require extensive parameter tuning to adapt to varying underwater conditions, yet they generally lack robustness and adaptability across different scenarios. In contrast, deep reinforcement learning (DRL) has proven to be an effective solution for target hunting \cite{s22155910} \cite{9723442}, boasting strong autonomous exploration capabilities. By continuously interacting with the environment, DRL enhances AUVs' behavior through iterative learning and adaptation. However, DRL approaches often overlook modeling interactions and coordination among agents, which limits their effectiveness in collaborative tasks like multi-AUV hunting.
To address these limitations, recent studies such as \cite{10298247} and \cite{jmse11071257} introduce multi-agent reinforcement learning (MARL) for collaborative AUV hunting, leveraging MARL’s potential to optimize the joint trajectories of hunter AUV formations. 
Despite these advances, existing MARL frameworks for target hunting primarily rely on online RL, resulting in low data utilization. In contrast, offline RL offers improved data efficiency, which trains using pre-collected datasets, addresses this limitation. Therefore, we propose an adaptive multi-agent diffusion policy (AMADP) algorithm. AMADP utilizes the powerful policy generation capabilities of diffusion models to model the trajectories of hunter AUVs and integrates an adaptive attention mechanism to dynamically adjust formations, improving coordination among hunter AUVs while maintaining covert communication constraints.
In summary, our contributions are as follows:
\begin{itemize}
    \item As far as we know, this is the first attempt to consider the eavesdropping capabilities of the target and propose a covert communication-guaranteed collaborative target hunting framework, enabling efficient coordination under covert communication constraints. 
    \item We design AMADP, a novel offline MARL algorithm,  which leverages diffusion models to model hunter AUV trajectories and utilizes adaptive attention to adjust hunter formation under complex constraints.
    \item Extensive experimental results demonstrate that the proposed AMADP algorithm outperforms current state-of-the-art MARL algorithms in terms of hunting success rate and convergence speed under covert communication constraints.
\end{itemize}
\begin{figure}[t]
\centerline{\includegraphics[scale=0.065]{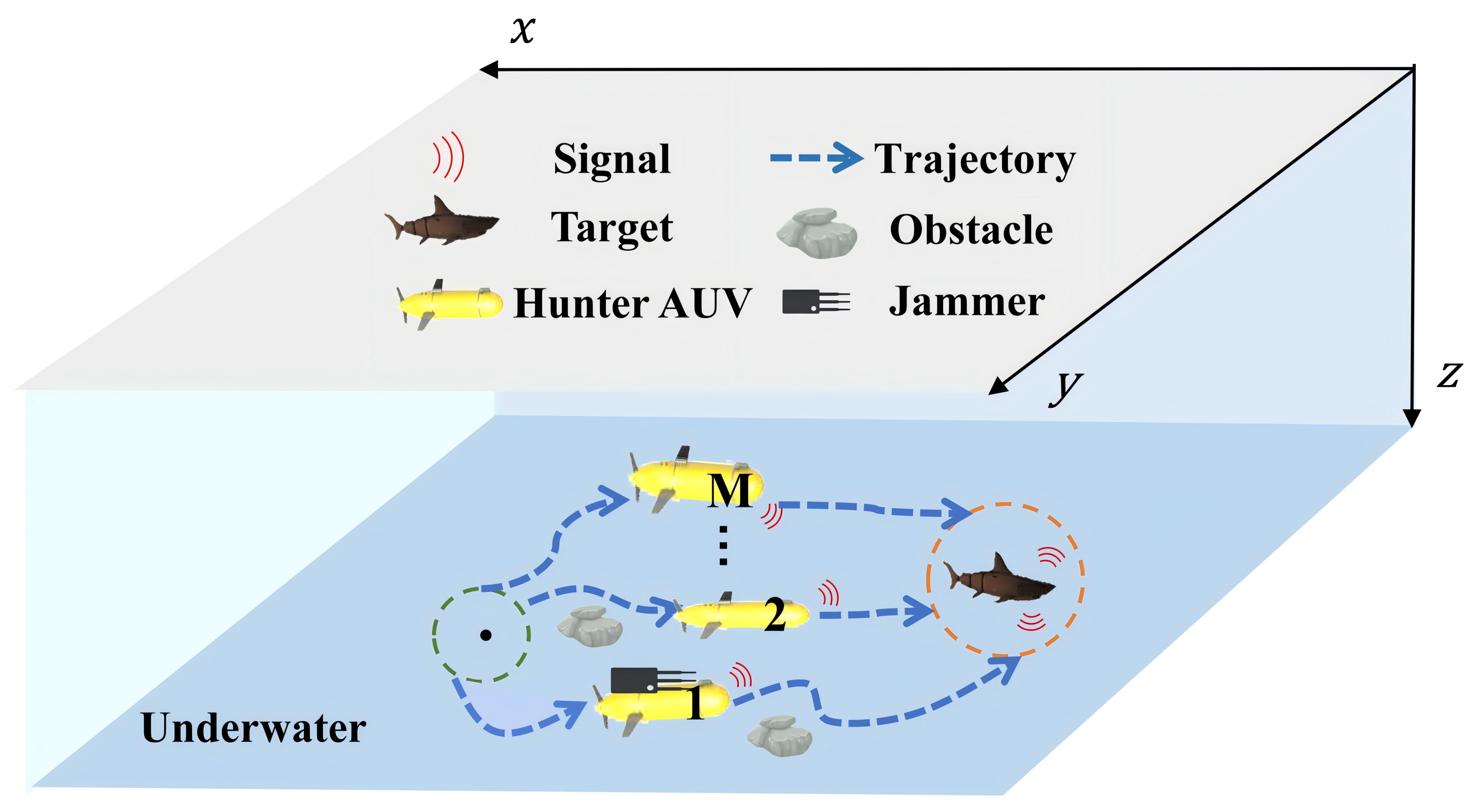}}
\caption{Illustration of the AUVs hunting scenario with covert communication.}
\label{model}
\end{figure}
\section{System Model and Problem Formulation}
The system model is illustrated in Fig.~\ref{model}. This scenario involves multiple hunter AUVs operating on a two-dimensional plane at a constant depth \( z \) underwater. The positions of hunter AUV \( i \) and the target at time \( t \) are represented by $\bm{L}_i(t) = (x_i(t), y_i(t), z)$ and $\bm{L}_{\mathrm{T}}(t) = (x_{\mathrm{T}}(t), y_{\mathrm{T}}(t), z)$, respectively. The hunter formation comprises \( M \) AUVs that collaboratively work to encircle  the target while avoiding underwater obstacles. Hunter AUVs communicate with each other using a fixed transmission power \( P_{\text{S}} \), exchanging critical information necessary for coordination. Furthermore, the first hunter AUV in the formation is equipped with a jammer that emits noise with power \( N_{\text{j}} \) to prevent the target from eavesdropping, ensuring covert communication throughout the hunting process.
\subsection{AUV Dynamics Model}

In the hunting scenarios considered, the dynamic models of AUVs can be expressed by a simplified three-degree-of-freedom model which describes the motion of the AUVs in the horizontal plane with a body-fixed coordinate frame 
\(
\bm{v_i} = [w_i, v_i, r_i]^T
\) and an earth-fixed reference frame 
\(
\bm{\eta_i} = [u_{xi}, u_{yi}, \psi_i]^T,
\)
where \( w_i \), \( v_i \), \( r_i \), and \( \psi_i \) represent the surge, sway, heave velocities, and yaw angle. $\bm{v}_i$ is constrained by the maximum limit \( V_1 \) satisfying \( \| \bm{v}_i \| \leq V_1 \). Similarly, we assume the velocity of the target $\bm{v}_\text{T}$ is bounded by \( V_2 \), i.e. \( \|\bm{v}_\text{T}\| \leq V_2 \). The hunter AUVs and target use the same dynamics model\cite{10298247}, which be given by
\begin{equation}
\begin{cases}
\dot{\bm{\eta}} = \bm{J}(\bm{\eta}) \bm{v}, \\[8pt]
\bm{M}_\text{A} \dot{\bm{v}} + \bm{C}_\text{A}(\bm{v}) \bm{v} + \bm{B}_\text{A}(\bm{v}) \bm{v} + \bm{G}_\text{A}(\bm{\eta}) = \bm{p} + \bm{e},
\end{cases}
\tag{1}
\end{equation}
where $\bm{M}_\text{A}$, $\bm{C}_\text{A}(\bm{v})$ and $\bm{B}_\text{A}(\bm{v})$ represent the inertia matrix including added mass, the Coriolis-centripetal force matrix, and the damping matrix of the AUV, respectively. Additionally, $\bm{G}_\text{A}(\bm{\eta})$ denotes the combined gravity and buoyancy matrix. The control input and environmental disturbance are represented by $\bm{p}$ and $\bm{e}$. Besides, $\bm{J}(\bm{\eta})$ is the transformation matrix, which can be given by
\[
\bm{J}(\bm{\eta}) =
\begin{bmatrix}
\cos \psi & -\sin \psi & 0 \\
\sin \psi & \cos \psi  & 0 \\
0          & 0            & 1
\end{bmatrix}. \tag{2}
\]

In the underwater target hunting task, hunter AUVs show better chasing ability by fulfilling cooperation, thus we assume the acceleration of hunter and the target satisfies \( \|\dot{\bm{v}}_i\| < \|\dot{\bm{v}}_\text{T}\| \). 

\subsection{Underwater Acoustic Channel Model}
We formulate the path loss in shallow water environments as 
\[
A(d,f) = d^m a(f)^d, \tag{3}
\]
where \(m\) represents the spreading factor characterizing the propagation geometry, \(f\) is the communication frequency, and \(d\) is the distance. \(a(f)\) denotes the attenuation coefficient per kilometer at a frequency \(f\) in kHz, is given by Thorp's empirical formula as
\small
\begin{align*}
10\lg a(f) &= 3.3 \times 10^{-3} 
+ \frac{0.11f^2}{1 + f^2} 
+ \frac{44f^2}{4100 + f^2} + 3.0 \times 10^{-4}f^2. \tag{4}
\end{align*}
\normalsize

Underwater environmental noise includes turbulence noise \(N_\text{t}\), shipping noise \(N_\text{s}\), wind-driven wave noise \(N_\text{w}\), and thermal noise \(N_\text{th}\). Based on the underwater noise model, the total noise level can be expressed as
\begingroup
\small
\begin{equation}
\left\{
\begin{array}{l}
10\lg N_\text{t}(f) = 17 - 30\lg f, \\[5pt]
10\lg N_\text{s}(f) = 40 + 20(s - 0.5) + 26\lg f - 60\lg(f + 0.03), \\[5pt]
10\lg N_\text{w}(f) = 50 + 7.5 \sqrt{w} + 20\lg f - 40\lg(f + 0.4), \\[5pt]
10\lg N_\text{th}(f) = -15 + 20\lg f,
\end{array}
\right.
\tag{5}
\end{equation}
\normalsize
\endgroup
where \(s\) and \(w\) denote shipping activity factor and wind speed. These noise sources are modeled as Gaussian processes, and the total underwater environment noise power spectral density is given by \(N_\text{u}(f) = N_\text{t}(f) + N_\text{s}(f) + N_\text{w}(f) + N_\text{th}(f)\) in dB re \(\mu\text{Pa}\) per Hz. The acoustic path loss and underwater noise power at \(t\)-th time step are denoted as \(A[t]\) and \(N_\text{u}[t]\). 
\subsection{Covert Communication Model}
 We assume that the channel state is constant during each time step but changes independently from one time step to the next. To describe the underwater time-varying channel state, we use a block fading channel model, considering each channel to be mutually independent. The hunter AUVs transmit signals through multiple channels during communication, represented as $\bm{y}^{\text{R}}[t] = [y_1^{\text{R}}[t], \ldots, y_L^{\text{R}}[t]]$, where $L$ denotes the number of channels uses. Meanwhile, the target passively monitors the signal strength to detect communication among hunter AUVs. The signal received by the target on the $l$-th channel in the $t$-th time step is expressed as\cite{10090449}
\begingroup
\setlength{\abovedisplayshortskip}{10pt}
\setlength{\belowdisplayshortskip}{10pt}
\begin{equation}
y_l^{\text{T}}[t] = 
\begin{cases} 
\sqrt{\frac{P_\text{S}[t]}{A_{\text{ae}}[t]}} s_l + n_l^{\text{T}}[t], & H_1, \\
n_l^{\text{T}}[t], & H_0, 
\end{cases}
\tag{6}
\end{equation}
\endgroup
where hypothesis $H_1$ represents that the hunter AUVs are communicating with each other, and $H_0$ indicates they are not. $P_\text{S}[t]$ and $s_l$ denote the transmit power and the transmitted signal on the $l$-th channel in the $t$-th time step, respectively. Besides, $n_l^{\text{T}}[t] \sim \mathcal{N}(0, N_{\text{T}}[t])$,  where $N_{\text{T}}[t] = N_\text{u}[t] + N_\text{j}[t]$, represents the total Gaussian noise observed by the target in the $t$-th time step and $A_\text{ae}[t]$ refers to the acoustic path loss experienced by the signal as it propagates from the hunter AUVs to the target.

In this scenario, the target infers the communication state between hunter AUVs by measuring the received signal power and applying a hypothesis testing strategy to aid its decision-making during the hunting task.
Specifically, to detect the presence of the covert communication, target needs to clarify whether a hunter AUV sends information to others. During the hunting process, the target employs the likelihood ratio test (LRT)\cite{steven1993fundamentals} as the optimal detection approach to determine whether the hunter AUVs are communicating, while minimizing its detection error. This process can be simplified as
\begin{equation}
Y[t] \triangleq \frac{1}{L} \sum_{l=1}^{L} \left| \tilde{y}_l^\text{T}[t] \right|^2 \underset{H_0}{\overset{H_1}{\gtrless}} \alpha[t], \tag{7}
\end{equation}
where $Y[t]$ represents the average received signal power of the target, $\sum_{l=1}^{L} \left| \tilde{y}_l^\text{T}[t] \right|^2$ is the total received power at the target over the block duration, and $\alpha[t]$ is the threshold for the $t$-th time slot. If the received power is below the threshold, the target favors hypothesis $H_0$; otherwise, $H_1$.

Under the assumption that target knows the transmission power $P_\text{S}[t]$ of the hunter AUVs and noise power $N_\text{T}[t]$, and the optimal threshold $\alpha^*[t]$ at target is given by\cite{10553242}
\vspace{5pt}
\begin{equation}
\alpha^*[t] = N_\text{T}[t] \left( 1 + \frac{1}{\beta_\text{T}[t]} \right) \ln(1 + \beta_\text{T}[t]), \tag{8}
\end{equation}
where $\beta_\text{T}[t]$ equals $\frac{P_\text{S}[t]}{A_\text{ae}[t] N_\text{T}[t]}$.
\vspace{5pt}

Let $P_\text{FA}$ represent the probability of false alarm, indicating that the target favors hypothesis $H_1$ when the actual state is $H_0$. Similarly, $P_\text{MD}$ denote the missed detection probability indicating a preference for $H_0$ when $H_1$ is true. 
To ensure covert communication, the following constraint must be satisfied
\begin{equation}
P_\text{FA} + P_\text{MD} \geq 1 - \epsilon, \tag{9}
\end{equation}
where $\epsilon$ signifies the acceptable level of covertness.
However, calculating $P_\text{FA}[t]$ and $P_\text{MD}[t]$ directly is challenging. Therefore, according to\cite{10553242}, the constraint equation can be transformed into
\begin{equation}
\frac{L}{2} \left[ \ln(1 + \beta_T[t]) - \frac{\beta_T[t]}{1 + \beta_T[t]} \right] < 2\epsilon^2. \tag{10}
\end{equation}

\subsection{Problem Formulation}
The proposed covert communication-guaranteed collaborative target hunting framework requires completing the hunting successfully while ensuring covert communication constraints throughout the entire process.

\textbf{Successful hunting criteria:}
We assume that the detection range and the attacking range of the hunter AUVs are $R_1$ and $R_2$, respectively. Specifically, when the distance between the target and the $i$-th hunter AUV $e_i$ is less than $R_1$ ($\|e_i\| < R_1$), the hunter AUVs obtain the target’s location information and share it within the formation. hunting is successful if all $M$ hunter AUVs are within the distance $R_2$ from the target ($\|e_i\| < R_2$) and form an encirclement. Conversely, if at the end of the given time steps $h$ all $\|e_i\| > R_1$, or if the conditions for a successful hunting are not met within this time, the hunting operation is considered a failure. 

\textbf{Covert communication constraint:}
The total divergence between the probability distributions under hypotheses $H_0$ and $H_1$ is constrained as
\begin{equation}
D_\text{KL}(Q_0 \| Q_1)[t] = \frac{L}{2} \left[ \ln(1 + \beta_\text{T}[t]) - \frac{\beta_\text{T}[t]}{1 + \beta_\text{T}[t]} \right]\leq 2 \epsilon^2, \tag{11}
\end{equation}
where $\epsilon$ represents the error tolerance, and $\mathcal{D}_\text{KL}\left(Q_0 \| Q_1 \right)[t]$ is the Kullback-Leibler (KL) divergence between the distributions under different hypotheses at time $t$. $Q_1$ and $Q_0$ represents the probability density under hypothesis $H_1$ and $H_0$, respectively.

\textbf{Remark 1:} \textit{For simplicity, we assume that the transmission power $P_\text{S}[t]$ and communication frequency $f$ between 
 hunter AUVs and the jammer power $N_\text{j}[t]$ directed at the target all remain constant. Therefore the covert communication constraints are solely dependent on the distances between the hunters and the target. Our goal is to optimize the trajectories of the hunter AUVs in the formation to complete the hunting task while satisfying the covert communication constraints.}

 In summary, we define the optimization problem as 
\begin{equation}
\begin{aligned}
\max \; & P_{\text{success}} = \Pr \left( \|e_i\| < R_2 \quad \forall i = 1, \dots, M \right) \\
\text{s.t.} \quad 
& \begin{cases}
    \quad D_\text{KL}(Q_0 \| Q_1)[t] \leq 2 \epsilon^2, & \forall t, \\[8pt]
    \quad \|\bm{v}_i(t)\| \leq V_1, \quad \|\bm{v}_{\mathrm{T}}(t)\| \leq V_2,  & \forall t, \\[8pt]
    \quad \|\bm{L}_i(t) - \bm{L}_j(t)\| \geq r_{\min}, & \forall i \neq j.
\end{cases}
\end{aligned}
\tag{12}
\end{equation}
where \( P_{\text{success}} \) represents the probability of successful encirclement when all hunter AUVs are within the target's capture range \( R_2 \). \( D_\text{KL}(Q_0 \| Q_1)[t] \) enforces covert communication constraints, \( V_1 \) and \( V_2 \) are the maximum speeds for the hunters and the target, respectively, and \( r_{\min} \) is the minimum safe distance to avoid collisions.
\begin{figure}[t]
\centerline{\includegraphics[scale=0.075]{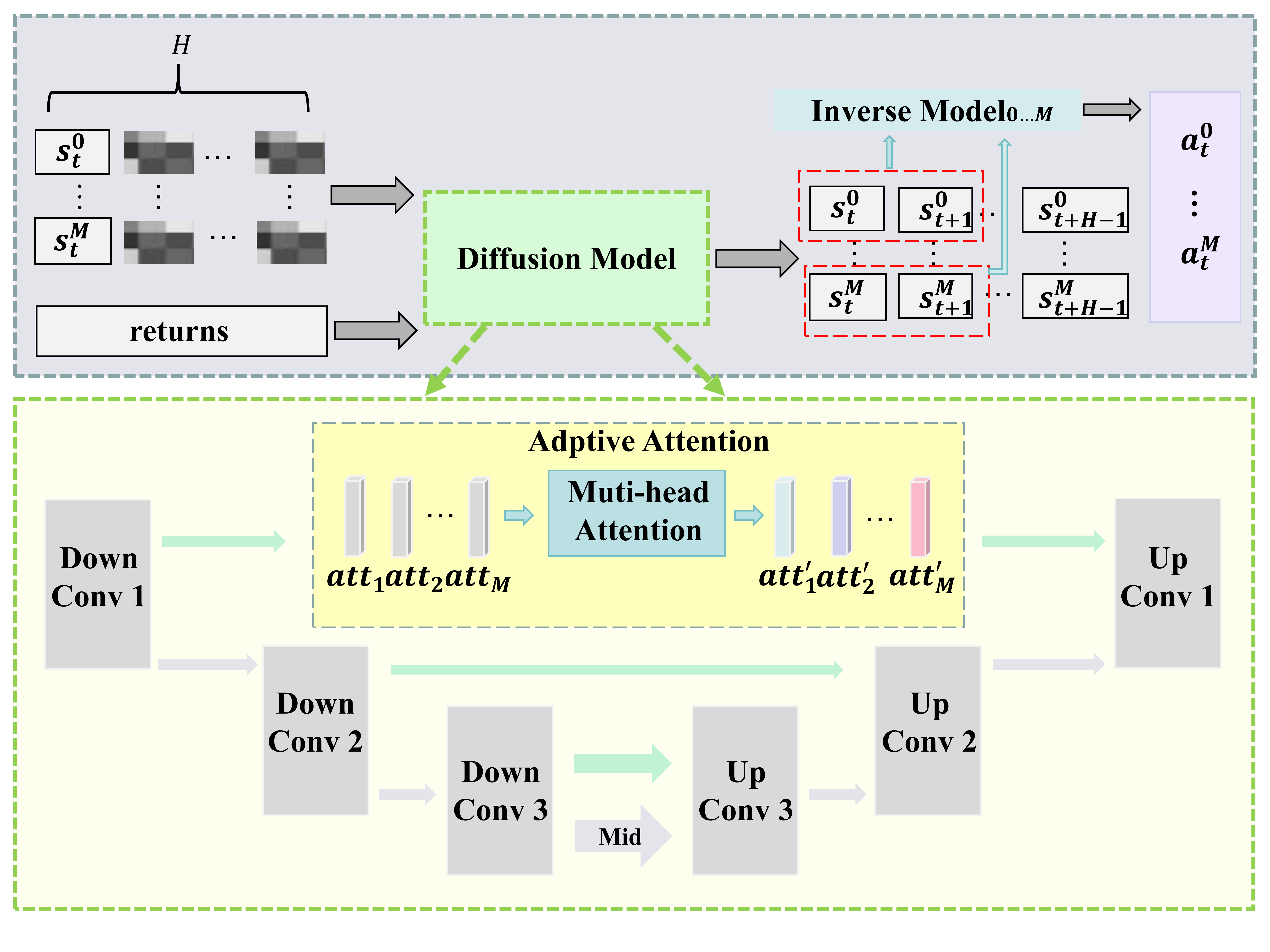}}
\caption{Illustration of the AUVs hunting scenario with covert communication.}
\label{algrithm}
\end{figure}
\section{Methodology}
\subsection{Markov Game Process Modeling}

We model the hunting process as a partially observable Markov decision process (POMDP), where each AUV and the target act as agents
\begin{equation}
M = (S_i, A_i, P_i, R_i), \tag{13}
\end{equation}
where \( S_i \) represents the observation information of each agent, \( A_i \) is the action taken by each agent, \( P_i(s'|s, a) \) determines the transition probability to the next state, and \( R_i \) is the reward information. In this context, hunter AUVs receive the same rewards.

\subsubsection{State Space}
The observation information for each agent can be defined as 
\begin{equation}
S_i(t) = \left\{ v_i(t), \, \bm{L}_i(t), \, \left\{ \bm{L}_\text{o}(t) \right\}, \, \left\{ \bm{L}_j(t) \right\}_{j \neq i} \right\}
, \tag{14}
\end{equation}
where $v_i(t)$ and $\bm{L}_i(t)$ represent the velocity and position information of agent $i$, respectively. $\left\{ \bm{L}_\text{o}(t) \right\}$ denotes the positions of the obstacles, and $\left\{ \bm{L}_j(t) \right\}_{j \neq i}$ represents the positions of other agents.
\subsubsection{Action Space}
Each agent's action \( a_i(t) \) includes the movement direction \( \theta_i \) and velocity \( v_i \), which is represented as
\begin{equation}
a_i(t) = \left\{ \theta_i(t), \, v_i(t) \right\}. \tag{15}
\end{equation}
\subsubsection{Reward Function}

To achieve effective encirclement and covert communication constraints under dynamic and uncertain underwater conditions, the reward function is designed to encourage the hunter AUVs to form an optimal encirclement around the target while maintaining specific communication constraints.

\begin{itemize} 
   \item \textbf{Encirclement formation reward \( R^\text{E}_i(t) \)}: This reward encourages hunters to form a well-distributed encirclement around the target,
\begin{equation}
R^\text{E}_i(t) = 
\begin{cases}
- \lambda \sigma_s(t), & d_g(t) > d_g^*, \\
\zeta (d_g^* - d_g(t)) - \lambda \sigma_s(t), & d_g(t) \leq d_g^*, 
\end{cases} 
\tag{16}
\end{equation}
where \( \sigma_s(t) \) represents the variance in the distances between the hunters to encourage a balanced formation, and \( d_g(t) \) is the distance between the target and the centroid of the hunter AUVs formation at time \( t \), and \( d_g^* \) is the desired distance for encirclement completion. When \( d_g(t) \leq d_g^* \), an additional reward is provided for a successful hunting task.

    \item \textbf{Collision avoidance reward \( R^\text{C}_i(t) \)}: 
    This reward penalizes collisions between hunter AUVs or with landmarks, helping maintain a safe distance during the hunting task.
\begin{equation}
R^\text{C}_i(t) = 
\begin{cases}
- \nu, & \text{if collision}, \\
0, & \text{otherwise}.
\end{cases}\tag{17}
\end{equation}

    \item \textbf{Covert performance reward \( R^\text{P}_i(t) \)}: 
    This reward is designed to ensure that the hunting process adhere to the covert communication constraint $\epsilon$. 
    \begin{equation}
    R^\text{P}_i(t) = 
    \begin{cases} 
    \nu, & D(Q_0 \| Q_1)[t] \leq 2\epsilon^2, \\
    -\nu, & D(Q_0 \| Q_1)[t] > 2\epsilon^2.
    \end{cases}\tag{18}
    \end{equation}
    
\end{itemize}

The overall reward for an hunter AUV \( R_i(t) \) is a  sum of the above rewards
\begin{equation}
R_i(t) =  R^\text{E}_i(t) +  R^\text{C}_i(t) +   R^\text{P}_i(t),\tag{19}
\end{equation}
This reward design ensures that hunters effectively perform hunting tasks while avoiding collisions and meeting covert communication requirements.

\subsection{Design of AMADP Algorithm}
\subsubsection{Diffusion Model}
The proposed algorithm is based on the denoising diffusion probabilistic model (DDPM) \cite{ho2020denoising}, a powerful deep generative model designed to learn the underlying data distribution \( q(\mathbf{x}_0) \) from a dataset \( \mathcal{D} = \{\mathbf{x}_k\} \). In DDPM, data reconstruction is achieved by denoising real data $\mathbf{x}_0$ from noises $\mathcal{N}(0, \mathbf{I})$ over $K$ diffusion steps.

The predefined forward process is defined as $q(\mathbf{x}_k|\mathbf{x}_{k-1}) = \mathcal{N}(\sqrt{\alpha_k}\mathbf{x}_{k-1}, \sqrt{1 - \alpha_k}\mathbf{I})$, where $\alpha_k = 1 - \beta_k$, and $\beta_{1:K}$ is a variance schedule. The reverse process is parameterized as $p_\theta(\mathbf{x}_{k-1}|\mathbf{x}_k) = \mathcal{N}(\mu_\theta(\mathbf{x}_k), \Sigma_k)$. 
The mean $\mu_\theta(\mathbf{x}_k)$ and variance $\Sigma_k$ can be expressed as $\mu_\theta(\mathbf{x}_k) = \frac{1}{\sqrt{\alpha_k}} \left( \mathbf{x}_k - \frac{\beta_k}{\sqrt{1 - \bar{\alpha}_k}} \epsilon_\theta(\mathbf{x}_k, k) \right)$ and $\Sigma_k = \beta_k \frac{1 - \bar{\alpha}_{k-1}}{1 - \bar{\alpha}_k} \mathbf{I}$, where $\bar{\alpha}_k = \prod_{i=1}^{k} \alpha_i$.

\enlargethispage{0.1in}
The loss function of DDPM is defined as 
\begin{equation}
L(\theta) = \mathbb{E}_{k \sim \{1,\ldots,K\}, \mathbf{x}_0, \epsilon} \left[\lVert \epsilon - \epsilon_\theta(\mathbf{x}_k, k) \rVert^2 \right],\tag{20}
\end{equation}
where $\epsilon$ is the real noise added in each diffusion step, and $\epsilon_\theta(\mathbf{x}_k, k)$ is the noise predicted by the noise prediction network at diffusion step \( k \).

\subsubsection{Architecture of AMADP}
The overall architecture of the network is illustrated in Fig.~\ref{algrithm}. The entire AUV formation shares a noise prediction network, which is based on a modified U-Net architecture consisting of three down-sampling convolutional layers and three up-sampling convolutional layers, connected by a middle bottleneck. The down-sampling layers progressively compress the state trajectories features while integrating the conditioning information. The up-sampling layers reconstruct the state trajectories by concatenating the intermediate features from the corresponding down-sampling layers via skip connections. After the modified U-Net predicts state trajectories for each hunter AUV. These trajectories are then fed into the respective inverse dynamics models, allowing each hunter to determine the necessary actions at each time step for optimal target hunting.

To coordinate the formation between multiple hunter AUVs, we introduce an adaptive attention mechanism between the initial down-sampling and the up-sampling layer of the U-Net. Unlike conventional self-attention mechanisms that focus solely on global information, the adaptive attention is designed to dynamically adjust attention weights based on both global and agent-specific inputs. Formally, adaptive attention can be expressed as
\begin{equation}
\left\{
\begin{array}{l}
att_i = \text{softmax} \left( \frac{Q_{i,t} \cdot K_{t}^T}{\sqrt{d_k}} \right) \cdot V_t, \\[8pt]
att = \text{concat}(att_1, att_2, \dots, att_M),
\end{array}
\right.
\tag{21}
\end{equation}
where \( Q_{i,t} \) is the query of the \(i\)-th AUV, \( K_t \) and \( V_t \) represent the global key and value information. \( att_i \) represents the attention weight for each AUV. The attention weight \( att \) reflects the global attention information, incorporating the contributions of each hunter during the hunting process. By dynamically adjusting these attention weights, the model can better capture the interactions between hunter AUVs and optimize their coordination for effective target hunting.


\subsubsection{Training Framework of AMADP}
Unlike online RL, which requires real-time interaction with the environment for continuous policy updates, offline RL relies on a pre-collected static dataset $\mathcal{D}$ to learn the policy, thereby improving data utilization.
Considering the challenges posed by obstacles and unstable communication in underwater environments make real-time interaction extremely challenging, AMADP adopts an offline RL training approach. Moreover, diffusion-based offline RL is suitable for solving cooperative games. Thus the hunting policy of the entire hunter AUV formation is generated by the AMADP algorithm proposed in this paper, and the escape policy of the target is the pre-trained traditional RL method deep deterministic policy gradient (DDPG)\cite{lillicrap2015continuous}.
 
In our approach, the diffusion model is conditioned on  \( y_{\tau} \), which contains the current state, achieved return, and the current time step to generate future trajectories. Considering the actual hunting process, AMADP uses a centralized training with decentralized execution (CTDE) framework, where global information is accessed during training, but each hunter AUV makes decisions based on local observations during execution.
To simplify the representation of the decision process in the diffusion model, we define the learning state sequence as
\begin{equation}
\tau = \left[ s_0^1, \ldots, s_0^M, \, s_1^1, \ldots, s_1^M, \, \ldots, \, s_H^1,\ldots, s_H^M \right],
\tag{22}
\end{equation}
where \( s_t^i \) indicates the state of the \(i\)-th hunter AUV at time step \(t\) and \(H\) represents the planning ahead time steps.

For each hunter AUV, we define an inverse dynamics model \( f_{\phi} \), which predicts the action \( a_t^i \) of the \(i\)-th AUV at time step \(t\) based on its current state \( s_t^i \) and next state \( s_{t+1}^i \)
\begin{equation}
a_t^i = f_{\phi} (s_t^i, s_{t+1}^i).\tag{23}
\end{equation}

Incorporating the DDPM loss, the inverse dynamics model loss, and the classifier-free guidance mechanism\cite{ho2022classifier}, the overall training loss function can be formulated as
\begin{align*}
\mathcal{L}(\theta, \phi) := &\sum_{i} \mathbb{E}_{(s_t^i, a_t^i, s_{t+1}^i) \in \mathcal{D}} 
\left[\left\lVert a_t^i - f_{\phi}^{i} (s_t^i, s_{t+1}^i) \right\rVert^2 \right] \\
&\kern-4em+ \mathbb{E}_{k, \tau \in \mathcal{D}, \beta} \left[
\lVert \epsilon - \epsilon_{\theta} 
( \hat{\tau}_k, (1 - \beta) y(\hat{\tau}) + \beta \emptyset, k) 
\rVert^2 \right],\tag{24}
\end{align*}
where \(\beta\) is sampled from a Bernoulli distribution to balance the conditioned and unconditioned diffusion process.
\vspace{0.06in} 
\begin{table}[t]
    \centering
    \renewcommand{\arraystretch}{1} 
    \setlength{\tabcolsep}{10pt} 
    \caption{Parameters of System and Algorithm} 
    \begin{tabular}{c|c}
        \hline
        \textbf{Parameters} & \textbf{Values} \\
        \hline
        Diffusion step ($K$) & 200 \\
        Learning rate & 0.0001 \\
        Training step & 20000\\
       
        Batch size & 32 \\
        Discounting factor  & 0.9 \\
        Return scale & 3000 \\
        Planning ahead time steps ($H$) & 40\\
        Reward design ($\lambda$, $\zeta$, $\nu$) & 20, 6000, 10 \\
        Start point of huner AUVs ($O$) & (500, 500, -200) m \\
        Number of hunter AUVs ($M$) & 3 \\
        Maximum speed of hunter AUV ($V_1$) & 0.3 m/s \\
        Maximum speed of target ($V_2$) & 0.2 m/s \\
        Acceleration of hunter AUV ($\|\dot{v}_{i}\|$) &\SI{0.01}{\meter/\second\squared} \\
        Acceleration of target  ($\|\dot{v}_\text{T}\|$) & \SI{0.02}{\meter/\second\squared} \\
        Movement range of AUV ($\psi$) & $[-\pi, \pi]$ \\
        Sensing radius of AUV ($R_1$) & 800 m \\
        Attacking radius of AUV ($R_2$) & 150 m \\
        Desired distance ($d_g^*$) & 120 m \\
       Communication Constraint ($\epsilon$) & 0.04 \\
       Communication Frequency ($f$) & 25 kHz \\
        Transmission Power ($P_\text{S}$) & 0.1 W \\
       Jammer Power ($N_\text{T}$) & 0.2 W \\
        \hline
    \end{tabular}
    \label{parameters}
\end{table}

\begin{figure*}[htbp] 
    \centering
    \begin{subfigure}{0.27\textwidth}
      \raisebox{-0.3cm}{ 
            \includegraphics[width=\linewidth]{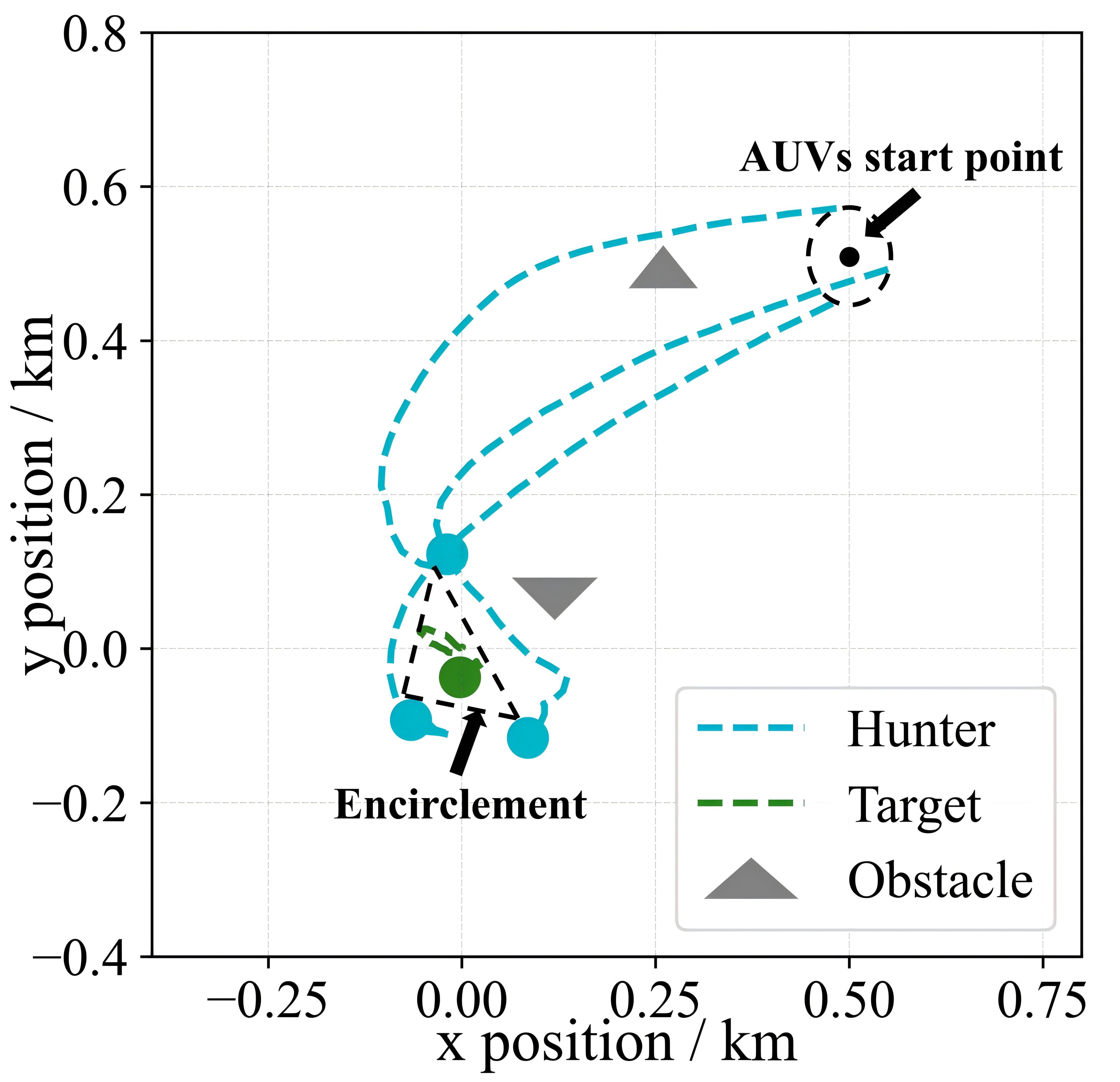}
        } 
        \caption{Hunting trajectory   with AMADP.}
        \label{fig:3a}
    \end{subfigure}
    \hfill
    \begin{subfigure}{0.32\textwidth}
       \raisebox{-1.2cm}{ \includegraphics[width=\linewidth]{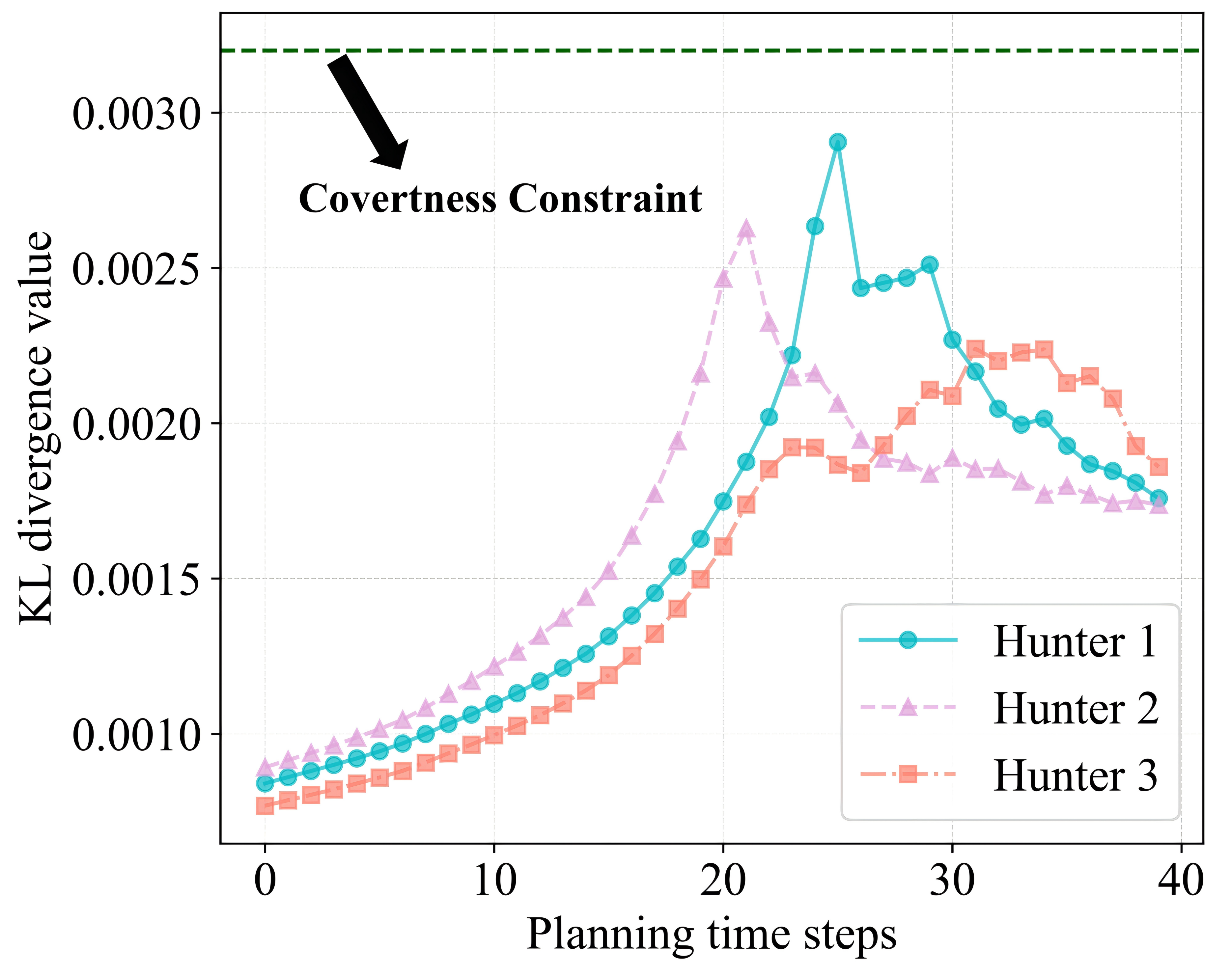}} 
       \caption{Covert  performance of AMADP.}
        \label{3b}
    \end{subfigure}
    \hfill
    \begin{subfigure}{0.34\textwidth}
        \raisebox{0.1cm}{\includegraphics[width=\linewidth]{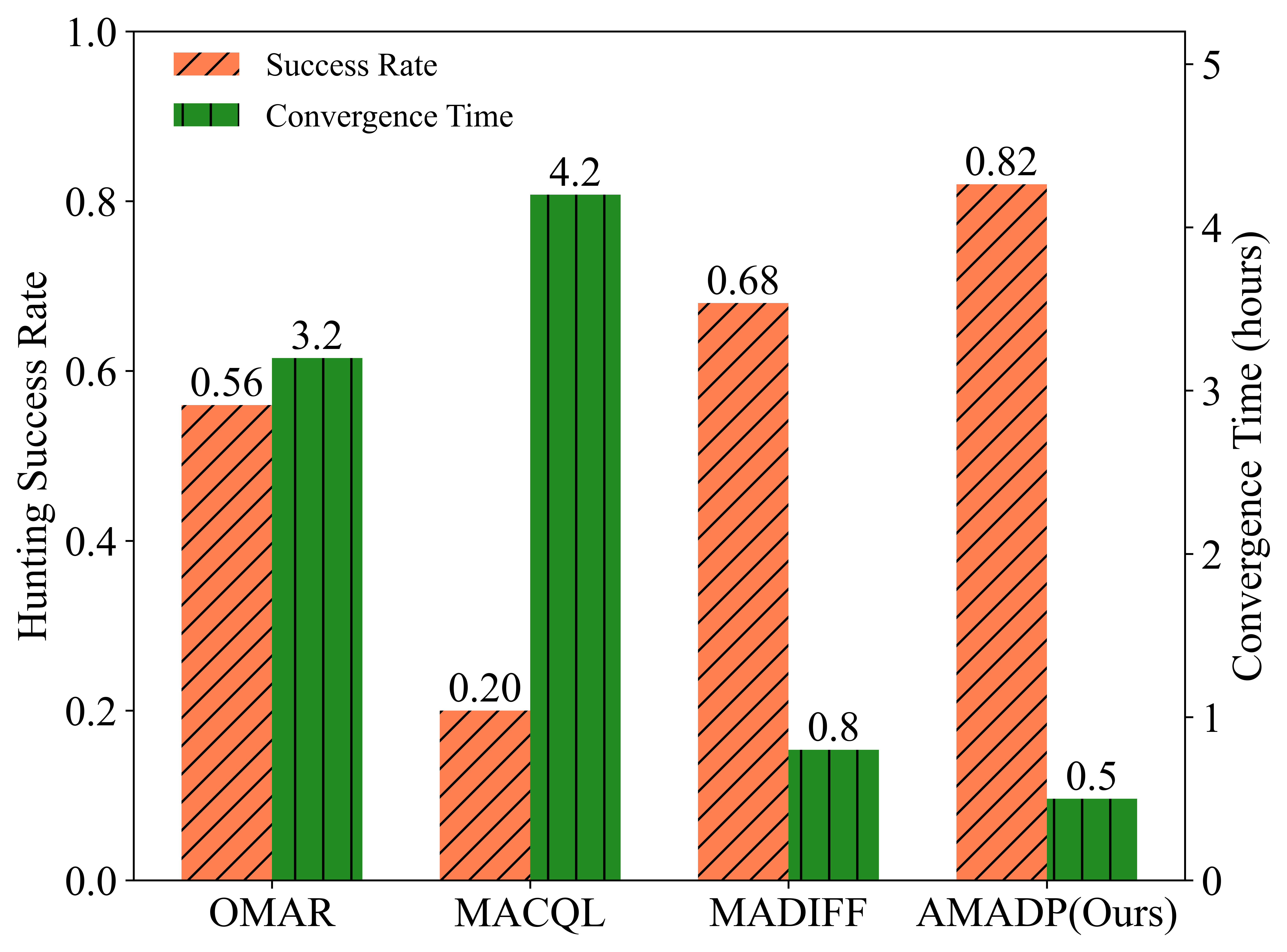}}
        \caption{ Hunting performance between algorithms.}
        \label{3c}
    \end{subfigure}
    \caption{Hunting with covert communication between hunter AUVs and the target.}
    \label{fig:combined}
\end{figure*}

\section{Experimental Results}
The experiments are conducted in a 1200 m × 1200 m area with a water depth of -200 m, where obstacles are randomly distributed. The AUV formation starts from the central point (500, 500) and the target's position is randomly initialized. The parameters of the system model and the AMADP algorithm are presented in Table~\ref{parameters}.

Fig.~\ref{fig:combined}(a) illustrates a successful encirclement using the AMADP algorithm, where the AUV formation successfully surrounds the target while avoiding obstacles without any collisions. We measured the $D_\text{KL}(Q_0 \| Q_1)$ values at each time step over 50 episodes, taking the average value of each time step, as shown in Fig.~\ref{fig:combined}(b). The results indicate that the AMADP algorithm consistently satisfies the covert communication constraints throughout the hunting process.

To demonstrate the superior performance of the proposed algorithm, we compare AMADP with state-of-the-art offline MARL algorithms, multi-agent conservative Q-learning (MACQL)\cite{kumar2020conservative}, OMAR\cite{pan2022plan} and MADIFF\cite{zhu2023madiff}. The experimental results in Fig.~\ref{fig:combined}(c) indicate that AMADP significantly outperforms other algorithms in the success rate of hunting task. And also due to the simplified structure of our proposed algorithm, it achieves faster convergence.

\section*{Conclusion}
In this paper, we consider the eavesdropping capabilities of the target during collaborative hunting and propose a covert communication-guaranteed hunting framework for the first time. To improve the generalization capability, data efficiency, and trajectory diversity of traditional collaborative hunting algorithms, we introduce AMADP, an offline 
 MARL algorithm, which incorporates diffusion models to generate diverse hunting trajectories and utilizes adaptive attention to dynamically adjust the formation. Experimental results demonstrate that AMADP satisfies covert communication constraints and achieves higher success rates and faster convergence compared to existing state-of-the-art algorithms.
\vspace{12pt}
\bibliographystyle{IEEEtran}
\bibliography{reference}  
\end{document}